\documentclass[aps,prl,twocolumn,showpacs,groupedaddress]{revtex4}

\usepackage{graphicx}

\begin{document}


\title{A Degenerate Bose-Fermi Mixture of Metastable Atoms}


\author{}

\author{J.\ M.\ McNamara}
\author{T.\ Jeltes}
\author{A.\ S.\ Tychkov}
\author{W.\ Hogervorst}
\author{W.\ Vassen}
\affiliation{Laser Centre Vrije Universiteit, De Boelelaan 1081, 1081 HV
 Amsterdam, the Netherlands}


\date{\today}

\begin{abstract}
We report the observation of simultaneous quantum degeneracy in a dilute gaseous Bose-Fermi mixture of metastable atoms. Sympathetic cooling of helium-3 (fermion) by helium-4 (boson), both in the lowest triplet state, allows us to produce ensembles containing more than $10^{6}$ atoms of each isotope at temperatures below 1~$\mu$K, and achieve a fermionic degeneracy parameter of $T/T_{F}=0.45$. Due to their high internal energy, the detection of individual metastable atoms with sub-nanosecond time resolution is possible, permitting the study of bosonic and fermionic quantum gases with unprecedented precision. This may lead to metastable helium becoming the mainstay of quantum atom optics.
\end{abstract}


\maketitle

The stable fermionic ($^3$He) and bosonic ($^4$He) isotopes of helium (in their ground state), as well as mixtures of the two, have long exhibited profound quantum properties in both the liquid and solid phases \cite{wilkj90}. More recently, the advent of laser cooling and trapping techniques heralded the production of Bose-Einstein condensates (BECs) \cite{ande95,davi95} and the observation of Fermi degeneracy \cite{dema99,trus01} in weakly interacting atomic gases. To date nine different atomic species have been Bose condensed, each exhibiting its own unique features besides many generic phenomena of importance to an increasing number of disciplines.

We can expect that studies of degenerate fermions will have a similar impact, and indeed they have been the object of much study in recent years, culminating in the detection of Bardeen-Cooper-Schriefer (BCS) pairs and their superfluidity \cite{zwie05}. However, only two fermions have so far been brought to degeneracy in the dilute gaseous phase: $^{40}$K \cite{dema99} and $^6$Li \cite{trus01}. Degenerate atomic Fermi gases have been difficult to realize for two reasons: firstly, evaporative cooling \cite{hess86} relies upon elastic rethermalizing collisions, which at the temperatures of interest ($<$~1~mK) are primarily s-wave in nature and are forbidden for identical fermions; and secondly, the number of fermionic isotopes suitable for laser cooling and trapping is small. Sympathetic cooling \cite{lars86,myat97} overcomes the limit to evaporative cooling by introducing a second component (spin-state, isotope or element) to the gas; thermalization between the two components then allows the mixture as a whole to be cooled.

In 2001 a BEC of helium atoms in the metastable $2\;^{3}\textup{S}_{1}$ state (He*) was realized \cite{robe01,pere01}; more recently we reported the production of a He* BEC containing a large ($>$~$10^7$) number of atoms \cite{tych06}. A quantum degenerate gas of He* is unique in that the internal energy of the atoms (19.8~eV) is many orders of magnitude larger than their thermal energy ($10^{-10}$~eV per atom at 1~$\mu$K), allowing efficient single atom detection with a high temporal and spatial resolution in the  plane of a microchannel plate (MCP) detector \cite{sche05}. In an unpolarized sample (as is the case in a magneto-optical trap (MOT)) the internal energy leads to large loss rates due to Penning ionization (PI) and associative ionization (AI) \cite{stas06}: \[\textup{He*}+\textup{He*}\rightarrow \textup{He}+\textup{He}^++e^- \quad ( \mbox{or}\quad \text{He}_{2}^{+}+e^-).\] These losses are forbidden by angular momentum conservation in a spin-polarized He* gas, in which all atoms have been transferred into a fully stretched magnetic substate. Spin-polarization suppresses ionizing losses by four orders of magnitude in the case of $^4$He* \cite{shly94,pere01}; it is only this suppression of the loss rate constant to an acceptable value of $\approx$~$10^{-14} \text{cm}^{3}/\text{s}$ \cite{tych06,pere01} that has allowed the production of a BEC in this system \cite{robe01,pere01,tych06}. It has been shown that for $^4$He* \cite{shly94} a very weak spin-dipole magnetic interaction can induce spin-flips and mediate PI/AI; far from being a hinderance however, the ions produced during these inelastic collisions can allow us to monitor losses non-destructively and in real-time
\cite{tych06,seid04}.

\begin{figure*}
\begin{center}
$\begin{array}{ccc}
\scalebox{0.6}{\includegraphics{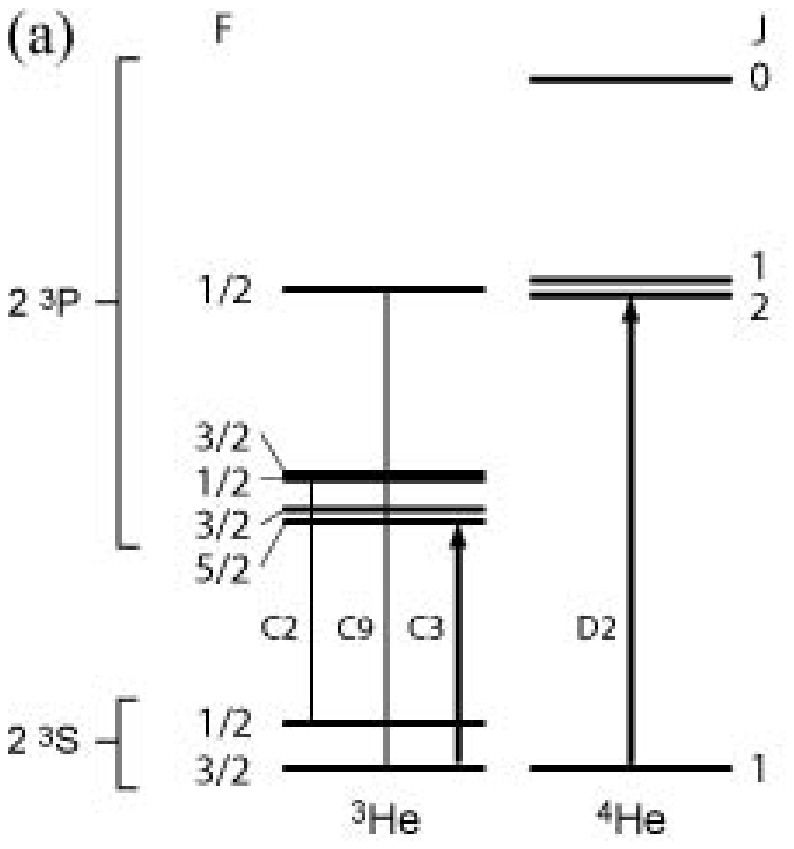}} & \scalebox{0.6}{\includegraphics{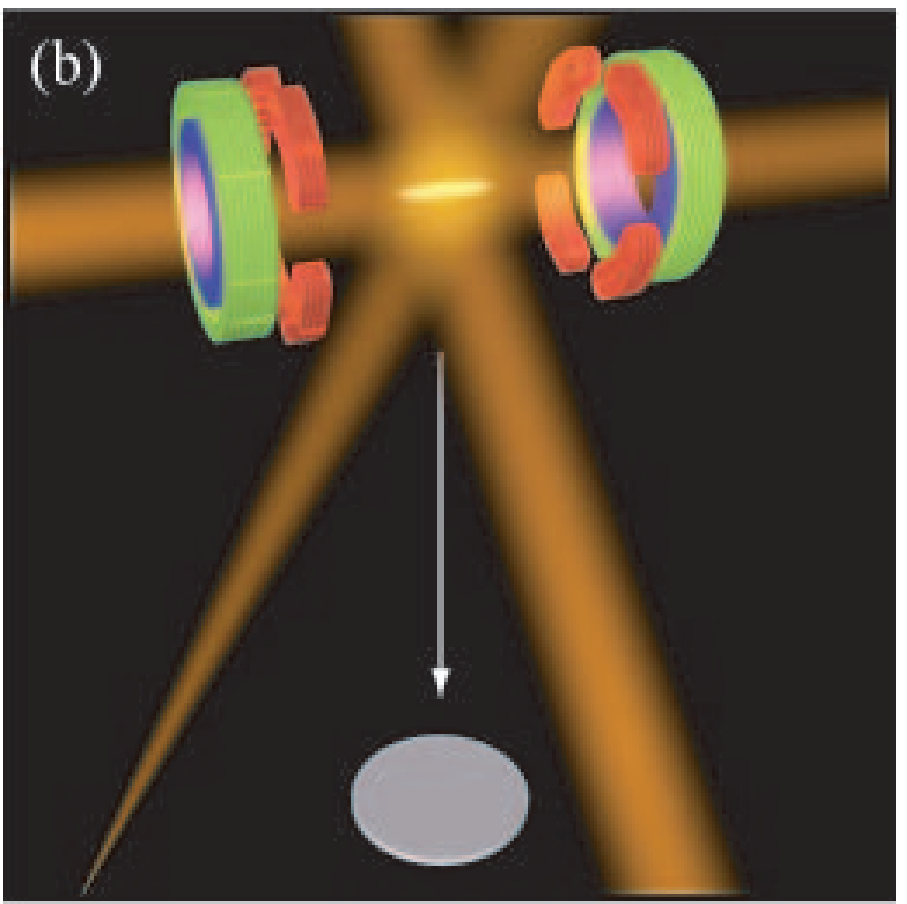}} & \scalebox{0.6}{\includegraphics{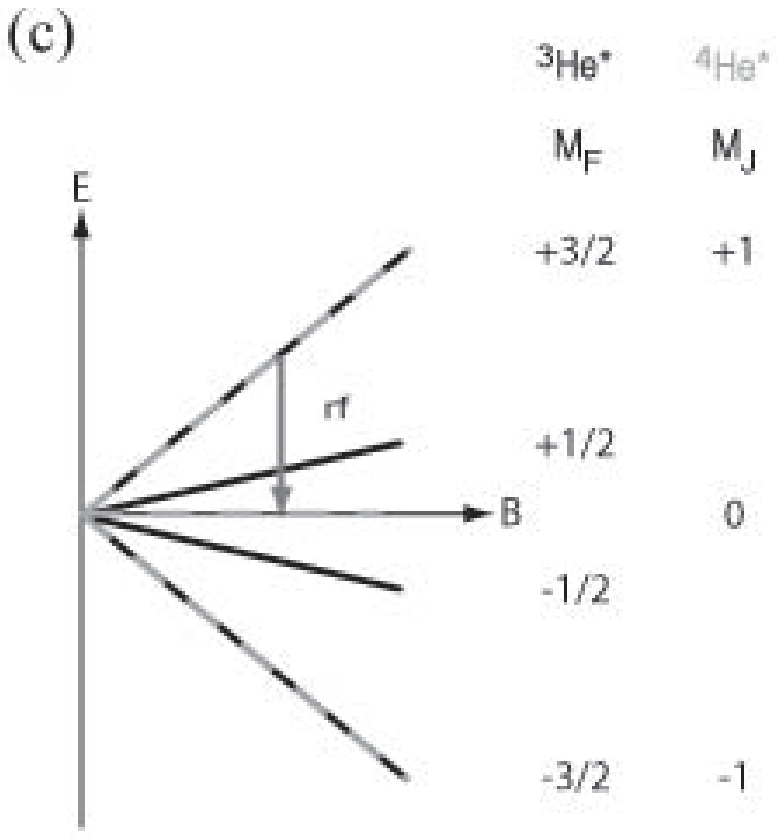}}
\end{array}$
\end{center}
\caption{{\bf (a)} $^3$He* and $^4$He* energy levels relevant to the magneto-optical trapping of He* atoms in the $2\;^{3}\text{S}_{1}$ state ($F$=3/2 for $^3$He*); the cycling transitions (C3 and D2) are indicated (MOT and Zeeman slower detunings are: $\delta_{\mathrm{MOT}}=-40$~MHz and $\delta_{\mathrm{ZS}}=-250$~MHz respectively). A repumper exciting the C2 transition ($\delta_{\mathrm{rp}}=-52$~MHz) is applied by simply double passing our $^3$He* Zeeman slower AOM. {\bf (b)} (color online). Experimental setup for the magneto-optical trapping, magnetic trapping and detection of $^3$He* and $^4$He* atoms. A TOF experiment is performed by dropping the trapped sample on an MCP detector positioned 17~cm below the trap center. {\bf (c)} Magnetic field dependence of the $^3$He* $F$=3/2 and $^4$He* $J$=1 magnetic substates. Evaporative cooling is performed on the $^4$He* $M_{J}=1\rightarrow {M_{J}}'=0$ transition.}
\label{fig1}
\end{figure*}

In $^3$He* the hyperfine interaction splits the $2\;^{3}\textup{S}_{1}$ state into an inverted doublet ($F$=3/2 and $F$=1/2, where $F$ is the total angular momentum quantum number) separated by 6.7~GHz (Fig.~\ref{fig1}a).
The only magnetically trappable state for which spin conservation should hold is the $|F,M_{F}\rangle =|3/2,+3/2\rangle$ substate (where $M_{F}$ is the projection of {\it F} on the quantization axis). Whether or not interactions would enhance spin-flip processes and the associated loss rate was unknown, and the prospect of a similarly acceptable level of suppression in the case of $^3$He* (and indeed between $^3$He* and $^4$He*) was an open question before this work.

Having realized the capability of producing large BECs of $^4$He* \cite{tych06}, and therefore large clouds of ultracold $^4$He*, we use $^4$He* to sympathetically cool a cloud of $^3$He* into the quantum degenerate regime. In the manner demonstrated previously \cite{stas04}, we have adapted our setup \cite{tych06} to allow the magneto-optical trapping of both He* isotopes simultaneously. The present configuration traps a mixture of $N_{^3\text{He*}}=7\times 10^8$ and $N_{^4\text{He*}}=1.5\times 10^9$ atoms simultaneously at a temperature of $\approx$~1~mK; a complication here is the need for a repumper exciting the $^3$He* C2 transition, due to the near (-811~MHz) coincidence of the $^4$He* laser cooling and $^3$He* C9 transitions (Fig.~\ref{fig1}a) \cite{stas04}. Unable to cool so much $^3$He* \cite{carr04}, we reduce the number of $^3$He* atoms in the two-isotope MOT (TIMOT) to $\approx$~$10^7$ by either altering the ratio $^3$He:$^4$He in our helium reservoir or, more simply, by loading the TIMOT with $^3$He* for a shorter period.

Spin-polarization of the mixture to the $^3$He* $|3/2,+3/2\rangle$ and $^4$He* $|1,+1\rangle$ states prior to magnetic trapping not only suppresses PI and AI, but also enhances the transfer efficiency of the mixture into the magnetic trap. The application of 1D-Doppler cooling along the symmetry axis of our magnetic trap \cite{tych06} (Fig.~\ref{fig1}b) reduces the sample temperature to $T=0.13$~mK without loss of atoms, increasing the $^4$He* phase space density by a factor of 600 to $\approx$~$10^{-4}$, greatly enhancing the initial conditions for evaporative cooling. We note at this point that the application of 1D-Doppler cooling to the $^4$He* component already leads to sympathetic cooling of $^3$He*, however the process appears to be more efficient if we actively cool both components simultaneously. During these experiments the lifetime of a pure sample of either $^3$He* or $^4$He* in the magnetic trap was limited by the background pressure in our ultra-high vacuum chamber to $\approx$~$110\;\mbox{s}$, whilst the lifetime of the mixture was only slightly shorter at $\approx$~$100\;\mbox{s}$, indicating that the suppression of PI and AI during $^3$He*-$^3$He* and $^3$He*-$^4$He* collisions works very well.

\begin{figure}
\scalebox{0.8}{\includegraphics*[viewport= 85 0 440 455,width=0.9\columnwidth]{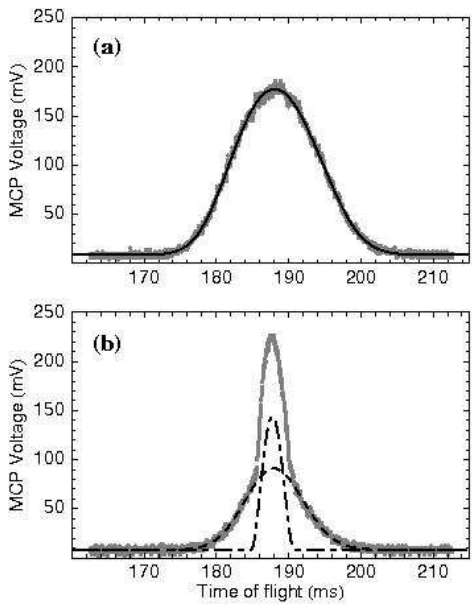}}
\caption{Time-of-flight spectra for: {\bf (a)} a degenerate Fermi gas of $^3$He* together with the fitted Fermi-Dirac TOF function, from which we conclude that $N_{^3\text{He*}}=2.1\times 10^{6}$ at $T=0.8$~$\mu$K and $T/T_{F}=0.45$; and {\bf (b)} a degenerate mixture of $N_{^3\text{He*}}=4.2\times 10^{5}$ ($T/T_{F}=0.5$) and $N_{^4\text{He*}}=1\times 10^{5}$ atoms. The dashed curve is the result of fitting a Fermi-Dirac velocity distribution to the wings, while the dot-dashed curve is a fit to a pure Bose-Einstein condensate (showing the characteristic inverted parabolic shape).}
\label{fig2}
\end{figure}

In order to further increase the collision rate in our cloud, we adiabatically compress it during 200 ms by increasing the trap frequencies to their final radial and axial values: $\nu_{r}=273$~Hz and $\nu_{a}=54$~Hz for $^3$He*, and $\nu_{r}=237$~Hz and $\nu_{a}=47$~Hz for $^4$He* (the difference is due to their differing masses). We now perform forced evaporative cooling on the $^4$He* component by driving radio-frequency (RF) transitions to the untrapped $M_{J}=0$ and -1 spin states (where $M_{J}$ is the projection of total electronic angular momentum quantum number J on the quantization axis), thereby sympathetically cooling $^3$He*. The atoms couple only weakly to the magnetic field, and the energies of the various magnetic sub-states vary linearly with magnetic field: $E_{M_{F/J}}=g \mu_{B} M_{F/J}B$, where $g$ is the gyromagnetic ratio, $\mu_{B}$ the Bohr magneton, and $B$ the magnetic field strength (Fig.~\ref{fig1}c). Because of the differing $^3$He* and $^4$He* gyromagnetic ratios (4/3 and 2 respectively) the frequency, at any given B-field, for transitions between the magnetic substates in $^4$He* is 3/2 times that of $^3$He* (Fig.~\ref{fig1}c) and we only remove $^4$He* during evaporative cooling (assuming that the mixture remains in thermal equilibrium). Furthermore, at the trap minimum ($B=3\;\mbox{G}$) the difference in transition frequencies is 2.8~MHz. Thus, when the temperature of the trapped sample is low enough ($<$~20~$\mu$K) we may selectively remove either $^3$He* or $^4$He* from the trap by applying an appropriate RF sweep (despite having to drive two transitions in order to transfer $^3$He* atoms into an untrapped magnetic substate). This allows us to perform measurements on the mixture as a whole, or on a single component. Upon release a time-of-flight (TOF) experiment is performed (Fig.~\ref{fig1}b); by fitting TOFs (Fig.~\ref{fig2}) with the applicable quantum statistical distribution functions we can extract the temperature of the gas and, having previously calibrated the MCP with absorption imaging \cite{tych06}, the number of atoms.

A single exponential ramp of the RF frequency to below 8.4~MHz removes all $^4$He* atoms from the trap, and leads to the production of a pure degenerate Fermi gas of $^3$He* (Fig.~\ref{fig2}a). An analysis of the TOF signals shows that we have achieved a maximum degeneracy ($N_{^3\text{He*}}=2.1\times 10^{6}$ at $T=0.8$~$\mu$K) of $T/T_{F}=0.45$ in a pure fermionic sample (Fig.~\ref{fig2}a), where the Fermi temperature is given by $k_{B}T_{F}=h(6N_{3}\nu_{r}^{2}\nu_{a})^{1/3}$, with $k_{B}$ Boltzmann's constant and $h$ Planck's constant. Alternatively we may halt the RF ramp just above 8.4~MHz and produce a quasi-pure BEC immersed in a degenerate Fermi gas (Fig.~\ref{fig2}b).

\begin{figure}
\scalebox{0.8}{\includegraphics*[viewport= 85 60 430 375,width=0.9\columnwidth]{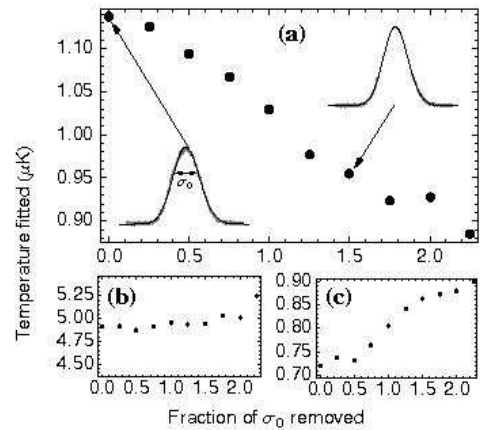}}
\caption{Temperatures obtained by repeatedly fitting 3 TOF spectra (from which increasingly large central fractions are removed) with a classical Maxwell-Boltzmann distribution: {\bf (a)} a degenerate $^3$He* Fermi gas ($T/T_{F}=0.5$), {\bf (b)} a thermal cloud of $^4$He* atoms with $T\gg T_{C}$ (a similar plot is observed for a Fermi gas with $T>T_{F}$) and {\bf (c)} a  Bose gas of $^4$He* atoms just above $T_{C}$, which adheres to Bose-Einstein  statistics.}
\label{fig3}
\end{figure}

Whilst recording a TOF we effectively integrate the density distribution of our sample over the two dimensions lying in the plane of our MCP detector, and the small difference between the non-gaussian distribution of our degenerate Fermi gas and the gaussian distribution of a classical gas becomes even less pronounced. It is therefore interesting to demonstrate the difference between the TOFs of classical gases and quantum gases explicitly, and confirm the result obtained above. As described by Schreck \cite{schreckthesis}, we repeatedly fit a gaussian distribution to a single TOF; before each fit a varying fraction of the TOF peak center is removed. The population of low energy states is suppressed (enhanced) in a cloud displaying Fermi-Dirac (Bose-Einstein) statistics, and fitting a gaussian distribution to the whole TOF will lead to an overestimation (underestimation) of the cloud size and therefore the temperature of the sample. By fitting only the more "classical" wings of a TOF these effects are negated, and the fitted temperature should either fall (fermions), rise (bosons), or stay constant in the case of a classical gas. The high signal-to-noise ratio of our TOF spectra allows us to see this behavior clearly (Fig.~\ref{fig3}).
By taking the temperature from a TOF for which we have removed $1.75~\sigma_{0}$ (where $\sigma_{0}$ is the root-mean-square width of a gaussian fit to the entire TOF, see Fig.~\ref{fig3}a), and the number of atoms calculated by integrating the TOF, we again recover a degeneracy parameter of $T/T_{F}=0.5$.

It is interesting to note that we have produced degenerate Fermi gases with evaporative cooling ramps as short as 2.5 s ($N_{^3\text{He*}}=4\times 10^{6}$ and $T/T_{F}=0.75$), signifying that the process of rethermalization is very efficient. At densities of $10^{10}-10^{12}$~atoms/cm$^3$ this indicates a large heteronuclear scattering length. Recently theory \cite{przy05} and experiment \cite{moal06} finally agreed upon the $^4$He*-$^4$He* scattering length ($a_{44}=7.64(20)$ and 7.512(5)~nm respectively). An extension to the theory of Przybytek and Jeziorski \cite{przy05} suggests that the $^3$He*-$^4$He* scattering length should also be very large and positive ($a_{34}=+28.8^{+3.9}_{-3.3}$~nm) \cite{przyprivcomm}. Such a large heteronuclear scattering length leads us to expect that losses, in particular boson-boson-fermion (BBF) 3-body losses (which scale with $a^{4}$), will have a significant impact on the mixture. We can estimate (order of magnitude) the BBF 3-body loss rate constant by using $K^{BBF}_{3}=120 \hbar a_{34}^{4}\sqrt{d+2/d}/(m_{4} \sqrt{3})$ \cite{inca04}, where $d$ is the $^3$He:$^4$He mass ratio, $m_{4}$ is the $^4$He mass and we assume the theoretical value of $a_{34}$ given above. This gives $K^{BBF}_{3}\approx 1.4\times 10^{-24}$~cm$^6$/s, indicating an atom loss rate that is 1-3 orders of magnitude larger than in the case of pure $^4$He*, and a condensate lifetime ($\tau_{C}$) which is significantly shorter in the degenerate mixture than in the absence of $^3$He*. These estimates are in agreement with initial observations that $\tau_{C}^{(3+4)}\sim 0.01$~s while $\tau_{C}^{(4)}\sim 1$~s \cite{tych06}. Given the large magnitude of $a_{34}$ and having seen no evidence for a collapse of the mixture, we may further suppose that $a_{34}$ is positive. This is then the first Bose-Fermi system to exhibit boson-fermion and boson-boson interactions which are both strong and repulsive. A possible disadvantage, however, may be that the system is only expected to be sufficiently stable against Penning ionization when the atoms are all in the fully stretched magnetic substates, hampering the exploitation of possible Feshbach or optical resonances.

In conclusion we have successfully produced a degenerate Fermi gas of metastable $^3$He containing a large number of atoms with $T/T_{F}=0.45$ and have also seen that we can produce a degenerate Bose-Fermi mixture of $^3$He* and $^4$He*. This source of degenerate metastable fermions, bosons, or mixtures of the two, could form the basis of many sensitive experiments in the area of quantum atom optics and quantum gases. Of particular interest is the recently realized Hanbury Brown and Twiss experiment on an ultracold gas of $^4$He* \cite{sche05}, demonstrating two-body correlations (bunching) with neutral atoms (bosons); we now have the ideal source to study anti-bunching in an ultracold Fermi gas of neutral atoms. The extremely large and positive $^3$He*-$^4$He* scattering length lends itself to the hitherto unobserved phenomena of phase separation in a Bose-Fermi mixture \cite{molm98} and, if the scattering lengths can be tuned only slightly, may allow a study of p-wave Cooper pairing of identical fermions mediated by density fluctuations in a bosonic component \cite{efre02}. Given the naturally large and positive scattering lengths, loading such a mixture into an optical lattice will provide a new playground for the study of exotic phases and phase transitions \cite{lewe04}, including supersolidity \cite{kim04}. The possibility of ion detection as a real-time, non-destructive density determination tool will be very helpful in observing these and other phenomena. Finally, the ultralow temperatures to which we can now cool both isotopes will allow unprecedented accuracy in high resolution spectroscopy of the helium atom. This could improve the accuracy to which the fine structure constant is known and may allow, via isotope shift measurements, an accurate measurement of the difference in charge radius of the $^3$He and $^4$He nucleus \cite{drak99}, challenging nuclear physics calculations.
\begin{acknowledgments}
We thank J. Bouma for technical support. This work was supported by the Space
 Research Organization Netherlands (SRON), Grant No. MG-051, the "Cold Atoms"
 program of the Dutch Foundation for Fundamental Research on Matter (FOM) and
 the European Union, Grant No. HPRN-CT-2000-00125.
\end{acknowledgments}

\bibliography{mcbib}

\end{document}